\input harvmac.tex
\voffset=-.61truein
\input epsf
\def\eps{\varepsilon}
\def\pr#1{{{#1}^\prime}}
\def\om{\omega}
\def\al{{\alpha}}
\def\dl{\delta}
\def\CM{{\cal M}}
\def\CN{{\cal M}}
\def\CV{{\cal V}}
\def\CZ{{\cal Z}}
\def\pa{\partial}
\def\eqdef{{\ {\buildrel{\scriptscriptstyle{\bf def}}\over=}\ }}%definici\'on
% first a full nine-point font set

%%%%%%%%%%%%%%%
\font\afw=msbm9
\font\af=msbm10 \font\afs=msbm8  \font\afbb=msbm16
 \font\bbm=bbm10
%\font\blackboard=msbm10
\font\blackboards=msbm7
\font\blackboardss=msbm5
\newfam\black
\textfont\black=\af \scriptfont\black=\blackboards
\scriptscriptfont\black=\blackboardss

%\def\spec#1{\hbox{\speci #1}}

%greek boldface
\catcode`\@=11               %% (read in bold math symbols)
\font\tenbmi=cmmib10 \skewchar\tenbmi='177 \font\sevenbmi=cmmib10
at 7pt \skewchar\sevenbmi='177 \font\fivebmi=cmmib10 at 5pt
\skewchar\fivebmi='177
\newfam\bmfam  
\textfont\bmfam=\tenbmi \scriptfont\bmfam=\sevenbmi
\scriptscriptfont\bmfam=\fivebmi

\mathchardef\bfA="7841 \mathchardef\bfb="7862
\mathchardef\bfX="7858 \mathchardef\bfY="7859

\mathchardef\bfalpha="780B \mathchardef\bfbeta="780C
\mathchardef\bfgamma="780D \mathchardef\bfdelta="780E
\mathchardef\bfepsilon="780F \mathchardef\bfzeta="7810
\mathchardef\bfeta="7811 \mathchardef\bftheta="7812
\mathchardef\bfiota="7813 \mathchardef\bfkappa="7814
\mathchardef\bflambda="7815 \mathchardef\bfmu="7816
\mathchardef\bfnu="7817 \mathchardef\bfxi="7818
\mathchardef\bfpi="7819 \mathchardef\bfrho="781A
\mathchardef\bfsigma="781B \mathchardef\bftau="781C
\mathchardef\bfupsilon="781D \mathchardef\bfphi="781E
\mathchardef\bfchi="781F \mathchardef\bfpsi="7820
\mathchardef\bfomega="7821 \mathchardef\bfvarepsilon="7822
\mathchardef\bfvartheta="7823 \mathchardef\bfvarpi="7824
\mathchardef\bfvarrho="7825 \mathchardef\bfvarsigma="7826
\mathchardef\bfvarphi="7827           %  (boldface greek \varphi)
\mathchardef\bfPhi="7809              %  (boldface greek \Phi)
\mathchardef\bfPsi="7808              %  (boldface greek \Psi)
\mathchardef\bfAlpha="7841 \mathchardef\bfBeta="7842
\mathchardef\bfGamma="7800 \mathchardef\bfDelta="7801
\mathchardef\bfEpsilon="7845 \mathchardef\bfZeta="785A
\mathchardef\bfEta="7848 \mathchardef\bfTheta="7802
\mathchardef\bfIota="7849 \mathchardef\bfKappa="784B
\mathchardef\bfLambda="7803 \mathchardef\bfMu="784D
\mathchardef\bfNu="784E \mathchardef\bfXi="7804
\mathchardef\bfPi="7805 \mathchardef\bfRho="7852
\mathchardef\bfSigma="7806 \mathchardef\bfTau="7854
\mathchardef\bfUpsilon="7807
%\mathchardef\bfPhi="7808
\mathchardef\bfChi="7843
%\mathchardef\bfPsi="7809
\mathchardef\bfOmega="780A
\catcode`\@=12
%new version
 \font\bmit=cmmib10            %Boldface math italic
\font\expo=cmmib10 at 10 true pt    %superscript version
   \newfam\boldmath 
    \textfont8=\bmit \scriptfont8=\expo \scriptscriptfont8=\expo
  \mathchardef\alpha="710B     \mathchardef\beta="710C
  \mathchardef\gamma="710D     \mathchardef\delta="710E
  \mathchardef\epsilon="710F   \mathchardef\zeta="7110
  \mathchardef\eta="7111       \mathchardef\theta="7112
  \mathchardef\kappa="7114     \mathchardef\lambda="7115
  \mathchardef\mu="7116        \mathchardef\nu="7117
  \mathchardef\xi="7118        \mathchardef\pi="7119
  \mathchardef\rho="711A       \mathchardef\sigma="711B
  \mathchardef\tau="711C       \mathchardef\phi="711E
  \mathchardef\omega="7121     \mathchardef\varepsilon="7122
  \mathchardef\varphi="7127    \mathchardef\imath="717B
  \mathchardef\jmath="717C     \mathchardef\ell="7160
  \mathchardef\partial"7140
%Greek CAPTIALS are already defined this way in plain.tex
%EXAMPLE: boldface alpha is {\bm\alpha}

%end of greek boldface

\hsize 11cm
\vsize 20cm
\Title
{\vbox{\baselineskip12pt\hbox{\phantom{lll}}\hbox{FSU TPI 01/06}
\hbox{DFTUZ--01.12}}} {\vbox{\centerline{Nahm Transform and Moduli
Spaces}
\medskip
\centerline {of {\afbb C}P$^N$-Models on the Torus}}}
\medskip
\centerline{\bf M. Aguado, M. Asorey}
\bigskip\centerline{{ Departamento de F\'{\i}sica Te\'orica.
Facultad de Ciencias}} \centerline{Universidad de Zaragoza.  50009
Zaragoza. Spain}\medskip
\centerline{and}
\medskip
\centerline{\bf A. Wipf}
\centerline{{ Theoretisch Physikalisches Institut, Universit{\"a}t Jena.}}
\centerline{{ Fr{\"o}belstieg
1, D-07743 Jena, Germany}}
\def\Z{{\af Z}}
\def\R{{\af R}}
\def\T{{\af T}}
\def\C{{\af C}}
\def\F{{\af F}}
\def\id{{\bbm 1}}

\def\bA{{\bf A}}

\def\ft#1#2{\textstyle{#1\over #2}}
\def\mtxt#1{\quad \hbox{{#1}}\quad}
\def\cpn{{\af C}P$^N$}
\def\zb{\bar{z}}
\def\wb{\bar{w}}
\def\exp#1{{{\rm e}^{#1}}}

\def\cpt{{\af C}P}
\def\cp{{\af C}P}
\def\thetafnn#1#2#3{\vartheta\left[\matrix{{#1}\cr{#2}}\right]({#3})}
\bigskip\bigskip
\centerline{\bf Abstract}
\bigskip
 There is a Nahm transform for two-dimensional gauge fields  
which establishes a one-to-one
correspondence between the orbit space of $U(N)$ gauge fields with
topological charge $k$ defined on a torus and that of $U(k)$ gauge
fields with charge $N$ on the dual torus. The main result of this
paper is to show that a similar duality transform cannot exist for
\cpn\ instantons. This fact establishes a significative difference
between 4-D gauge theories and \cpn\ models. The result follows
from the global analysis of the moduli space of  instantons based
on a complete and explicit parametrization of all self-dual
solutions on the two-dimensional torus. The boundary of the space
of regular instantons is shown to coincide with the space of
singular instantons.  This identification provides a new approach
to analyzing the role of overlapping instantons in the infrared
sector of \cpn sigma models.

\Date{}

\newsec{Introduction}

Two-dimensional sigma models have long been used as a testing
ground for a variety of ideas in non-perturbative Quantum Field
Theory, especially because of some remarkable similarities with
non-Abelian gauge theories in $3\!+\!1$ dimensions
\ref\Polyakov{A.M.~Polyakov, Phys.~Lett. {\bf 59B} (1975) 79.}
(for a review emphasizing this connection, see
\ref\NSVZ{V.~A.~Novikov, M.~A.~Shifman, A.~I.~Vainshtein,
V.~I.~Zakharov, Phys.~Rept. {\bf 116} (1984) 103.}).
%

%\noindent
They are scale invariant at the classical level and
asymptotically free at the quantum level, some possess topological
winding numbers, instantons, a chiral anomaly when coupled to
fermions and generate a dynamical mass by non-perturbative effects
at zero temperature and a thermal mass $\sim g^2 T$ at finite
temperature. In this respect the $O(3)$ nonlinear $\sigma$-model
with action
\eqn\othree{ S={1\over 8g^2}\int d^2x\, (\partial_\mu
\vec{n}\cdot
\partial_\mu\vec{n}),\qquad
\vec{n}^{\,2}(x)=1,} has been very extensively studied, specially
by its numerous interesting applications to condensed matter:
(anti) ferromagnetism, Hall effect, Kondo effect, etc (see Ref.
\ref\tsv{A.M. Tsvelik, {\it Quantum Field Theory in Condensed
Matter}, Cambridge U. Press, Cambridge (1995).} for a review
stressing this point of view). The model has also been used to
analyze the sphaleron induced fermion-number violation at high
temperature \ref\wimo{E. Mottola, A. Wipf, Phys.~Rev.~{\bf D39}
(1989) 588.}. By setting ${\vec n}=\Psi^\dagger
\vec{\sigma}\,\Psi$ with a normalized $\Psi\in\,$\C$^2$, the
action \othree\ can be rewritten in the equivalent form
\eqn\cpoform{ S = {1\over 2g^2} \int {\rm d}^2x \, \left| D_\mu
\Psi \right|^2,\mtxt{with} \vert\Psi(x)\vert=1\mtxt{and}
A_\mu=-i\Psi^\dagger \pa_\mu\Psi.}
It is invariant under gauge transformations
$\Psi\to\exp{i\lambda(x)}\Psi$ and hence $\Psi(x)$ may be viewed
as an element in \cp$^1$.

There are two natural generalizations of the $O(3)\sim
\hbox{\cp}^1$ model: \noindent  $O(N\!>\!3)$ models with action
\othree\ but with $\vec{n}\in S^{N-1}$ and
%
%\ref{\BZJ}{E.~Br\'ezin, J.~Zinn-Justin,
%Phys.~Rev.~{\bf B14} (1976) 3110.}{W.~A.~Bardeen, B.~W.~Lee,
%R.~E.~Shrock, Phys.~Rev.~{\bf D14} (1976) 985.}
%
 \cpn models with action \cpoform, but with $\Psi\in\,$\cpn \
instead of \cp$^1$
\ref\Eichenherr{H.~Eichenherr, Nucl.~Phys.~{\bf B146} (1978) 215.}
\ref\GoloPer{V.~L.~Golo, A.~M.~Perelomov, Phys.~Lett.~{\bf 79B}
(1978) 112.}.
In contrast to the $O(N)$ models they possess instanton solutions
for all $N,$ and a $\theta$ term can be added to the action so
that their topological properties can be explored
\ref\WittenonCPN{E.~Witten, Nucl.~Phys.~{\bf B149} (1979) 285.}.
These models are expandable in $1/N$ and have been solved in the
large $N$ limit \ref\DALDV{A.~D'Adda, M.~L\"uscher, P.~Di
Vecchia, Nucl.~Phys.~{\bf B146} (1978) 63.}. The role of
instantons and related sphalerons \ref\snippe{J. Snippe, Phys.
Lett. {\bf B335} (1994) 395-402} is crucial for physical effects
at $\theta\neq 0$ \ref\paf{M. Asorey, F. Falceto, Phys. Rev. Lett.
 {\bf 80}
(1998) 234-237}. In this paper we shall mainly be interested in
the structure of spaces of instantons and hence shall only
consider the \cpn\-extensions of the $O(3)$-model.

In particular we shall focus on the search of a variant of the
Nahm transform for $2$-dimensional models. In $4$-dimensions this
remarkable duality transformation relates different instanton
moduli spaces of gauge theories formulated on the four-torus
\ref\Nahmthree{W.~Nahm, Phys.~Lett.~{\bf 90B} (1980) 413.}.
More explicitly, it transforms a charge $k$ self-dual (instanton)
$SU(N)$ gauge potential $A$ on {\af T}$^4$ into a charge $N$
self-dual $U(k)$ potential ${\hat A}$ on the {\it dual} torus
$\hat{\hbox{\af T}}\,^4$ as follows:
\eqn\defNahm{ ({\hat A}_\mu)_{ij}(u) = -i \int_{\hbox{\afs
T}^4}d^4x\; \psi_i^{u\dagger}(x\,) {\partial\over{\partial
u^\mu}} \psi^u_j(x\,), }
$\{\psi^u_{j},j=1\dots,k\}$ being  $k$ orthonormal zero-modes of
the Dirac operator with shifted potential $A_\mu + 2\pi
u_\mu{\hbox{\id}}$, where the constant piece $u$ parametrizes the
dual torus. This transformation being involutive means that the
moduli space $\CM_k^N$ of $SU(N)$ instantons with charge $k$ 
is equivalent to  $\CM_N^k$ that of $SU(k)$ instantons with charge $N$
.

Our search for a corresponding Nahm transform for \cpn models on
the two-torus was motivated by the observation that the complex
dimension of the moduli space for charge $k$ instantons in \cpn \
is
\eqn\dimmodspace{ {\rm dim}\ {\cal M}_k^N = k(N+1), \qquad k>1,}
exhibiting a duality that may be conjectured to hold at the level
of moduli spaces, ${\cal M}^N_k \approx {\cal M}^{k-1}_{N+1}$.
This conjecture was further prompted by the fact that there are
no charge $1$ instantons on the $2$-torus for any value of $N$
 \ref\rr{J.~L.~Richard, A.~Rouet,
Nucl.~Phys.~{\bf B211} (1983) 447.}\ref\sp{J.~M.~Speight,
Commun.~Math.~Phys. {\bf 194} (1998) 513\semi
J.~M.~Speight, {\tt hep-th/0105142.}}, a property shared
with gauge theories on the $4$-torus
\ref\mmuk{S. Mukai, Nagoya Math. J. {\bf 81} (1981) 153}
\ref\BvB{P.~J.~Braam, P.~van Baal, Commun.~Math.~Phys.~{\bf 122}
(1989) 122\semi
H. Schenk, Commun.~Math.~Phys.~{\bf 116} (1988) 177.}.
Similarly as for gauge theories this would be a
consequence of such a duality, since there is no \cp$^0$
instanton. If this duality  exists the dynamics of the \cpn \
models should simplify in sectors with large $k$, as it happens
for large $N$.

The aim of the paper is to analyze the existence or not of a
generalized Nahm transform for these sigma models. Hence
we shall only analyze the instantons on a torus, the only
Riemann surface whose dual (Jacobian) is also a torus.  This kind
of compactification of space time corresponds to the choice of
periodic boundary conditions which are appropriate for the study
of  finite temperature effects \wimo.

The space time compactification also presents some technical
advantages. The action of an instanton does not depend on the
parameters of moduli space. This then leads to zero-modes of the
fluctuation operator in the instanton background. One expects
that for each parameter in moduli space there is one associated
zero-mode or that the number of zero-modes is not smaller than the
dimension of the moduli space. This expectation is not fulfilled
for the sigma models on \R$^2$: if one varies some moduli
parameter of the instanton one finds non-normalizable  zero-modes
\ref\Ward{R.~S.~Ward, Phys.~Lett.~{\bf 158B} (1085) 424.}. In a
compact space this can never happen, thus in our case both
methods of counting the dimension of moduli spaces of instantons
are equivalent.

Since \cpn \ spaces admit a K\"ahler structure,
$2$-dimensional \cpn-models can be extended to $\CN\!=\!2$
supersymmetric theories. More general purely bosonic or
supersymmetric nonlinear sigma models with K\"ahler target spaces
have been studied on topologically trivial space-times by several
authors, see
\ref\Ruback{
B.~Zumino, Phys.~Lett.~{\bf 87 B} (1979) 203\semi
L.~Alvarez-Gaum\'e, D.~Z.~Freeman, Commun.~Math.~Phys. {\bf 80}
(1981) 443\semi
P.~J.~Ruback, Commun.~Math.~Phys.~{\bf 116} (1988)
645\semi
J.~P.~Gauntlett, Nucl.~Phys.~{\bf B400} (1993) 103.}.
All these models admit regular instanton solutions, the
topological charge of which appears as lower BPS-bound on the
action.

In addition to the regular instanton solutions these models
possess singular ones. Although they are usually ignored, it has
been pointed out recently that these singular configurations may
be of relevance for some topological field theories
\ref\lns{A.~Losev, N.~Nekrasov, S.~Shatashvili,
Class.~Quant.~Grav.~{\bf 17} (2000) 1181.}.
In particular, their contribution to the renormalization group
flow of supersymmetric theories and correlation  functions of
topological invariants seems to be crucial.

We shall analyze singular instantons as boundary configurations of
moduli spaces of regular instantons. In particular, we shall
discuss if they can appear as  limit case of strongly overlapping
regular instantons as it happens for 4-dimensional Yang-Mills
instantons \ref\pm{M.~G.~P\'erez, T.~G.~Kovacs, P.~van Baal, {\tt
hep-ph/0006155.}}.

This paper is organized as follows: In the next section we
briefly recall the basic features of classical $2$-dimensional
\cpn-models and their instanton solutions. Section $3$ contains a
detailed analysis of  zero-modes of the associated Weyl operator
on the $2$-dimensional torus. In the following section we study
the zero-modes for the shifted gauge potential and explicitly
construct the Nahm transform for two-dimensional gauge fields. In
particular, we show how this transformation maps $U(1)$ gauge
fields with charge $k$ over a $2$-dimensional torus into $U(k)$
gauge fields with charge $1$ over the dual torus. Section $5$
deals with the global structure of \cpn\ moduli spaces of
instantons on {\af T}$^2$. A general method for their
construction is proposed. It is based on properties of the fiber
bundles associated with the $U(1)$ connection, and  yields  a
complete description of the moduli spaces ${\cal M}_k^N.$ Some
simple examples are worked out explicitly. The relevant
topological properties of these spaces are investigated in
section $6$. The main result is that no invertible Nahm transform
for regular instantons can exist since the corresponding moduli
spaces are topologically distinct. In section $7$ we investigate
the role of singular instantons as limiting cases of  strongly
overlapping regular instantons. A summary of our results and
conclusions are contained in  section $8$.

\newsec{Instantons in {\af C}P$^N$ models}
\def\zb{{\bar z}}
The classical {\af C}P$^N$ model in 2 Euclidean space-time
dimensions is defined by the action
\eqn\actioncpn{ S =
{N\over{2g^2}} \int {\rm d}^2x \, \left| D_\mu \Psi
\right|^2,\qquad D_\mu\Psi=\big(\partial_\mu-iA_\mu\big)\Psi,}
where $\Psi(x)=(\Psi_a(x)),\ a=0,\ldots,N $ is a
$(N\!+\!1)$-component complex field with values in \cp$^{N+1}$. We
consider $\Psi(x)$ to be normalized
\eqn\constraintcpn{
|\Psi(x)|^2 = \sum_{a=0}^N |\Psi_a(x)|^2 = 1 }
and configurations differing by a phase factor are identified,
\eqn\identifcpn{
\Psi(x) \sim {\rm e}^{i\alpha(x)}\,\Psi(x).}
The action includes a covariant derivative
$D_\mu = \partial_\mu - i A_\mu$ with respect to the composite
$U(1)$ gauge field defined in terms of
the sigma field by
\eqn\gpot{
A_\mu(x) := -i \Psi^\dagger \partial_\mu \Psi.}
With this gauge field the symmetry
\identifcpn\ can be restated  as a usual $U(1)$ gauge
invariance. Indeed, the action \actioncpn\
is invariant under the
phase transformation \identifcpn\ of $\Psi(x)$
if at the same time the composite field $A_\mu$
transforms as a true $U(1)$ connection,
\eqn\transfgaugea{
A_\mu \rightarrow A_\mu + \partial_\mu \alpha(x),}
as it follows from its very definition.
Actually, one may view $\Psi$ and $A_\mu$ in \actioncpn\
as independent fields. The algebraic equations of motion for
the gauge potential are then just \gpot.

A topological charge (instanton number) can be defined
\eqn\topolcharge{
k={1\over{2\pi}} \int {\rm d}^2x \, F_{12}
= {1\over{2\pi i}} \int {\rm d}^2x \, \varepsilon_{\mu\nu}
\partial_\mu \Psi^\dagger \partial_\nu \Psi
= {1\over{8\pi i}} \int {\rm d}^2x \, \varepsilon_{\mu\nu} D_\mu
\Psi^\dagger D_\nu \Psi .}
This charge takes integer values for smooth  configurations with finite
action, and thus the space of configurations splits  into
disconnected instanton sectors.

\noindent
Application of the Cauchy-Schwartz inequality to $D_\mu
\Psi,$ $i\varepsilon_{\mu\nu} D_\nu \Psi$ yields
\eqn\csineq{
S \geq {{N\pi}\over{g^2}} |k|, }
The minimal action is reached by  solutions
of the first order  equations
\eqn\cauchyriemanncov{
D_\mu \Psi(x) = \mp i \varepsilon_{\mu\nu} D_\nu \Psi(x). }
Antiselfdual solutions correspond to instantons  ($-$sign and
$k>0$), and
selfdual solutions to anti-instantons ($+$sign and $k<0$).

In  complex coordinates $z = x^1\!+\!i x^2$ the
(anti)selfdual equations \cauchyriemanncov\ can be written as
follows
\eqn\covholomorphy{\eqalign{
D_\zb \Psi &= (\partial_\zb - i A_\zb) \Psi = 0
\qquad {\rm instantons}\cr
D_z \Psi &= (\partial_z - i A_z) \Psi = 0
\qquad {\rm anti-instantons}, }}
where the complex components of the gauge field are
$A_z=\half(A_1 - i A_2),\ A_\zb = \half(A_1 + i A_2).$

This provides the first characterization of the solutions as
holomorphic solutions with respect to the holomorphic bundle
structure induced by the gauge field $A$.

In summary, up to some common normalization factor the components
of instanton field solutions are holomorphic sections of a line
bundle over space-time. In the plane there is an infinite
dimensional space of solutions, but only the constants have a
finite action and $k=0$. In the torus {\af T}$^2$ the different
holomorphic structures  are parametrized for fixed $k$ by the
points of the dual torus $\hat{\hbox{\af T}}\,^2$ and the space
of instantons has a finite dimension \rr \sp.

The topological charge of such a solution is the sum of
multiplicities of the zeros of any non-trivial component of
$\Psi.$ From now on we concentrate on instanton solutions
($k>0$). The anti-instanton case is analogous.

\newsec{\cpn-models on the torus and Fermionic zero-modes}

We may view the torus as \R$^2$ modulo a two-dimensional lattice
$\Lambda$ generated by two vectors $e_1$ and $e_2$. For
simplicity we will restrict to orthonormal vectors $e_\mu$ and
use dimensionless coordinates. In the sector with instanton
number $k$ we shall choose as transition functions $U_\mu$
relating the fields at $x$ and $x+e_\mu$, \eqn\bc{
\Psi_a(x+e_\mu)=U_\mu(x)\Psi_a(x)\quad,\quad A(x+e_\mu)=A(x)-i
U^\dagger_\mu(x) dU_\mu(x),} the gauge transformations
\eqn\transitionf{ U_1=e^{i\pi kx^2}\mtxt{and} U_2=e^{-i\pi k
x^1}.}

This means that $A$ is defined on a non-trivial line bundle
$E_k(${\af T}$^{\,2},$\C). In two dimensions  any gauge field $A$
induces a holomorphic bundle structure on $E_k$ (in four
dimensions this only holds for self-dual gauge fields). Hence
there is a local complex gauge transformation $h$ such that
\eqn\hstr{ A_\zb=ih\,\partial_{\zb} \,h^{-1}\mtxt{and}
D_{\zb}=\pa_{\zb}+h\, \pa_{\zb}\,h^{-1},} which trivializes the
connection $A$ (see e.g. Ref. \ref\afl{M. Asorey, F. Falceto, G.
Luz\'on, Contemp. Math. {\bf 219} (1998) 1}).

Next we construct and discuss the zero-modes of the Dirac operator
on the $2$-dimensional Euclidean torus. By the index theorem the
number of right-handed minus the number of left-handed zero-modes of
the Dirac equation,
$$ \Dsl_A\psi=0 $$
 depends  only on the first
Chern class of the gauge field. Since there are only zero-modes of
one chirality the total number of such modes in the fundamental
representation is $k$. Since they have definite chirality they
are completely determined by one non-trivial component in the
Weyl basis, i.e. by one ordinary complex function $\psi$ which we
will identify with the spinor field $\psi$ itself.

Hence, in complex coordinates a zero-mode solves the Weyl equation
\eqn\zm{ \left(\partial_{\zb} -i A_{\zb}\right)\psi=0,\qquad k>
0}
and must satisfy the same boundary condition \bc\
as the components of $\Psi$. Thus the fermionic zero-modes
fulfill the same differential equation and boundary conditions as
the components of the sigma field $\Psi$.

After trivializing the connection as in
\hstr\ the Weyl equation becomes a holomorphic condition,
$$
D_\zb\,\psi=0\Longleftrightarrow
\partial_\zb \,\chi=0,\mtxt{where}\psi=h\chi.$$

\noindent However, as pointed out earlier, if $k\neq 0$ the
transformation $h$ cannot be globally defined and this shows up
in the the change of boundary conditions between $\psi$ and
$\chi$, $$ \chi(z+1)=\tilde U_1(z)\chi(z)\quad,\quad
\chi(z+i)=\tilde U_2(z)\chi(z),$$ where $$ \tilde
U_\mu(z)=h^{-1}(x+e_\mu)\,U_\mu(x)\, h(x) $$ is holomorphic. The
holomorphic character of the $\tilde U_\mu$ also reflects the
fact that any holomorphic section $\chi$ defines the holomorphic
structure of the bundle $E_k$  associated to the gauge field $A$.

Now we consider the particular gauge potential
\eqn\abel{ A^I_1 = -\pi kx^2,\qquad  A^I_2 = \pi kx^1,
\mtxt{or}  A^I_\zb = \ft{i}{2}\pi k z,}
which gives rise to a constant field strength
$F_{01}=2\pi k$ and instanton number $k$.
The complex gauge transformation trivializing
$A^I$ reads
\eqn\hgt{
h=e^{-\pi kz\zb/2}}
and the $\chi$ satisfy the holomorphic
boundary conditions with transition functions
\eqn\nbc{
\tilde U_1 = \exp{(1+2 z)\pi k/2}\mtxt{and}
\tilde U_2 = \exp{(1-2i z)\pi k/2}. }
The zero-modes can be conveniently expressed in terms of
Jacobi's theta functions
\eqn\th{\eqalign{
\thetafnn {a}{b}{\tau} &= \sum_{n=-\infty}^\infty
\exp{i\pi \tau (n+a)^2}
\exp{2\pi i(n+a)b}\cr
&=\eta(i\tau)\,\exp{2\pi i ab} q^{{a^2\over 2}-{1\over 24}}
\prod_{n>0}\big(1+q^{n+a-{1\over 2}}\exp{2\pi i b}\big)
\big(1+q^{n-a-{1\over 2}}\exp{-2\pi i b}\big),}}
where we have set $q=\exp{\,2\pi i \tau}$. These holomorphic and
quasi-periodic functions have the following shift properties
\eqn\thetaperiod{ \thetafnn {a\!+\!m\!+\!{n\over\tau}}{b}{\tau}=
\exp{2\pi in (a+b/\tau +n/2\tau)}\, \thetafnn {a}{b}{\tau}} and
possess first order zeros at the points
\eqn\zeroestheta{ \tau
a+b\in\{m+\ft12 +\tau(n+\ft12),\qquad m,n\in \hbox{\Z}\}.}
In terms of these $\theta$-functions a basis of linearly
independent zero-modes reads
\eqn\zerotheta{
\psi_\ell(x) =(2k)^{1\over 4}h(x)\,\chi_\ell(z),\qquad
\chi_\ell(z)=\exp{\pi k
z^2/2} \thetafnn{ z\!+\!{\ell\over k}} {0} {ik}}
and $\psi_\ell$
has $k$ zeros at the following points on the torus:
\eqn\zeroes{
x^1=\big\langle \ft12-\ft{\ell}{k}\big\rangle\quad,\quad
x^2_p=
\big\langle\ft{1}{k}(\ft12+p)\big\rangle,\qquad
p=1,2,\dots,k,}
where $\langle a\rangle$ denotes the unique element in the lattice
$\{a+\hbox{\af Z}\}$ lying in $[0,1)$.
%class of numbers in $a+$\Z.
The basis $\{\psi_\ell\}$ is orthonormal, $$
(\psi_\ell,\psi_{\ell'}) =\int_{\hbox{\afs T}^2}
\psi_\ell^\ast\psi_{\ell'}=\delta_{\ell\ell'}.$$
Let $x_{\ell p}\in\hbox{\af T}^2,\;p=1,\dots k,$ be the $k$ zeros
of $\psi_\ell$. Then their sum is independent of $\ell$ and is
given by \eqn\sumrule{ \sum_{p=1}^k x_{\ell p}=\big\langle
\ft{k}{2}\big\rangle\,e\quad{\rm modulo}\quad\Lambda,
\mtxt{where}e=e_1+e_2.} That (for fixed $k$) the sum of the zeros
is the same for all $\psi_\ell$ (modulo the lattice defining the
torus), follows from the fact that all holomorphic sections
$\chi_l(z)$  satisfy the same boundary conditions. This statement
holds true for any choice of a zero-mode basis.

Under a translation by $1/k$ in either of the two directions on
the torus the space of zero-modes is left invariant. This is
expected on general grounds and is needed for the Nahm transform.
More explicitly, let ${\bfpsi}$ denote the $k$-dimensional column
vector with entries $\left(\psi_1,\dots,\psi_k\right)$. Then the
transformations read
\eqn\zerotrans{ \bfpsi(x+\ft1k
e_1)=\exp{i\pi x^2}S_1\bfpsi(x)\mtxt{and} \bfpsi(x+\ft1k
e_2)=\exp{-i\pi x^1}S_2\bfpsi(x),}
with unitary $k\times k$ matrices
\eqn\shift{ S_1= \pmatrix{0&1&&\cr &\ddots&\ddots&\cr
&&0&1\cr 1&&0&0}\mtxt{and} S_2=\pmatrix{\al&0&&0\cr 0&\al^2&&0\cr
&&\ddots&\cr 0&0&&\al^k},\quad \al=\exp{-2\pi i/k},}
satisfying
\eqn\cocycle{ S_1^k=S_2^k=\hbox{\id}
%\hbox{\af I}
\mtxt{and} S_1S_2=\al
S_2S_1.}
These shift identities are consistent with the position of the
zeros of $\psi_\ell$ given in \zeroes. Note that $\bfpsi$ may be
viewed as a zero mode of the $U(k)$ potential $A^I$\id \ on the
smaller torus with circumferences $1/k$ and with instanton number
$1$. The last relation in \cocycle\ just guarantees that $$
\exp{i\pi x^2}S_1\mtxt{and} e^{-i\pi x^1}S_2$$ in \zerotrans\ are
consistent $U(k)$ transition functions on the smaller torus, that
is, they satisfy the cocycle conditions with periods $1/k$.

We could as well have taken an alternative set of orthonormal
zero modes,
\eqn\alternative{ \tilde\bfpsi (x)=S \bfpsi
(x),\qquad S^\dagger S=\hbox{\id}.}
For example using the zero-modes
\eqn\zerothetaalt{ \tilde\psi_\ell(x) =(2k)^{1\over
4}\,\exp{-\pi k z\,(\zb+z)/2} \thetafnn{ iz\!+\!{\ell\over k}}
{0} {ik}}
instead of the ones in \zerotheta\ amounts to
exchanging $x^1$ and $x^2$ in the formulae above. With respect to
the new basis one again finds the shift identity \zerotrans\ with
the replacements $$ S_\mu\longrightarrow S\,S_\mu\,S^{-1}.$$ The
algebraic relations \cocycle\ remain intact and hence are
independent of the choice of basis.

\newsec{Nahm transform of gauge fields on $2$-dimensional torus {\af T}$^2$}

Let $A$ be an arbitrary two-dimensional U(N) gauge field with
topological charge
$$ c_1(A)=\int_{\hbox{\af T}^2}\ \tr\,
F_{12}(A)$$
defined on a torus {\af T}$^2$.
Its Nahm transform is defined in terms of the zero-modes of the
Weyl operator for a shifted vector potential
\eqn\nnabel{ A^u_\mu= A^I_\mu+2\pi u_\mu\hbox{\id}}
which has the same topological charge.
By the index theorem the dimension
of the space of zero-modes of $\Dsl_{A^u}$ is $k$.
Let $\{\psi^u_j; j=1,2,\cdots k\}$ be an orthonormal basis of
zero-modes.
The Nahm transform assigns to the $U(N)$ gauge potential $A$ a
$U(k)$ potential $\hat{A}$ on the {\it dual} torus
$\hat{\hbox{\af T}}\,^2$ with topological charge $N$ as follows:
\eqn\naNahm{ ({\hat A}_\mu)_{ij}(u) = -i \int_{\hbox{\afs
T}^2}d^2x\; \psi_i^{u\dagger}(x\,) {\partial\over{\partial
u^\mu}} \psi^u_j(x\,), }
Note that this construction does not require any special
constraint on the original gauge field $A$ as it does
in four dimensions where $A$ must be selfdual.
This is because any 2-dimensional
gauge field defines a holomorphic structure in the corresponding
bundle, whereas in four dimensions this is true only for self-dual
gauge fields.

%\nref\adhm{M.~F.~Atiyah, V.~G.~Drinfeld, N.~J.~Hitchin,
%Yu.~I.~Manin, Phys.~Lett.~{\bf 65A} (1978) 185.}

%
%\newsec{Nahm transform for Abelian gauge fields}

Sigma model fields are associated to Abelian gauge fields. But
the Nahm transform does not preserve the Abelian character as we
shall see below. This already is the first indication that 
it might be problematic to extend the Nahm transform to sigma
models. To analyze this problem let us now apply the Nahm
construction to the Abelian field \abel.

An orthonormal basis of the zero-modes of the
Dirac equation for a
shifted vector potential
\eqn\nabel{ A^u_\mu= A^I_\mu+2\pi u_\mu\mtxt{or}
A^u_\zb=\ft{i}{2}\pi kz+\pi w,\qquad w=u^1+iu^2,}
can be constructed from the solutions
with $u=0$ by shifting the arguments
\ref\sw{I.~Sachs, A.~Wipf, Helv.~Phys.~Acta {\bf 65} (1992) 652\semi
%I.~Sachs, A.~Wipf, Ann. of Physics {\bf 249} (1996) 380\semi
U.G.~Mitreuter, J.M.~Pawlowski, A.~Wipf,
Nucl. Phys. {\bf B514} (1998) 381\semi
H.~Joos, Helv. Physica Acta {\bf 63} (1990) 670\semi
H.~Joos, S. Azakov, Helv. Phys. Acta {\bf 67} (1994) 723.}
\eqn\szm{
\psi^u_\ell(x)=\exp{i\pi(u,x)}\;
\psi_\ell\big(x+\ft{1}{k}\eps\,u\big),\qquad
\eps=\pmatrix{0&1\cr -1&0}.}
For later purposes it is useful to discuss some properties of
these zero-modes:

The $k$ zeros of these modes are related to those of the
$\psi_\ell$ by the shift in \szm, \eqn\zerosu{ x^1_p\in\big\langle
\ft12-\ft{1}{k}(\ell+u^2)\big\rangle\mtxt{and} x^2_p\in\big\langle
\ft{1}{k}(\ft12 +u^1 +p)\big\rangle,\qquad p=1,\dots,k.} Hence
two different modes share no common zero unless $u\in\,$\Z$^2$.

From \szm\ and \zerotrans\ one sees at once
that the vector $\bfpsi^u$ transforms in the same way as $\bfpsi$ when
either $x^1$ or $x^2$ is translated by $1/k$,
\eqn\zerotransu{ \bfpsi^u(x+\ft1k e_1)=\exp{i\pi
x^2}S_1\bfpsi^u(x)\mtxt{and} \bfpsi^u(x+\ft1k
e_2)=\exp{-i\pi x^1}S_2\bfpsi^u(x),}
where the matrices
$S_\mu$ have been introduced in \shift\ .

The $k$-dimensional subspace spanned by the zero modes is also
left invariant by the following simultaneous rotations of $x$
and $u$:
\eqn\rot{ (x,u)\longrightarrow
(\eps^nx,\eps^nu),\quad n\in\{0,1,2,3\}.}
This can be seen by
checking that the transformed states satisfy the same
differential equation and boundary condition as the original
ones. These rotations are represented by unitary $k\times k$
matrices on the subspace spanned by the zero-modes. They
are a remnant of the rotation symmetry on the torus 
for constant field strength.

In addition, the idempotent transformation
\eqn\idem{
(x,u)\longrightarrow (x^\prime,u^\prime)
=(\ft{1}{k}\eps u,-k\eps x)}
is projectively represented on
the eigenmodes
\eqn\projrep{ \bfpsi^{u^\prime} (x^\prime)=
\exp{-2\pi i (x,u)}\,\bfpsi^u(x).}

Under simultaneous translations of $x$ and $u$ the zero modes
are invariant, up to a phase
\eqn\trans{ \bfpsi^{u+\al}\big(x)=
\exp{i\pi (\al,\,x+{1\over k}\eps x)}
\bfpsi^u\big(x+\ft1k\eps\al\big).}
Later in this paper this
shift identity will be rather important.

Finally note, that for $u\neq 0$ the
holomorphic gauge transformation \hgt\ does not
trivialize the gauge field $A^u$ anymore.
The modified trivializing transformation reads
$$h^u(x)=
\exp{-{\pi\over 2}(kz\zb\,+{1\over k}w\wb-2i\, \zb w)}
,\qquad w=u^1+iu^2.$$
It not only transforms the unitary basis \szm\ into
a $z$-holomorphic basis but also into an $w$-holomorphic basis,
\eqn\holobasis{
\psi^u_\ell=(2k)^{1\over 4}\,h^u(x)\,\chi_\ell^w(z),\qquad
\chi^w_\ell=\chi_\ell(z-\ft{i}{k}w),}
where the $\chi_\ell$ have been introduced in \zerotheta.
This is the essential feature of the Nahm transform.
It follows that the  Nahm transformed gauge field ${\hat A}=
({\hat A}_{\ell\pr\ell})$, defined by the Mukai-Nahm construction,
\eqn\nahmtrans{
({\hat A}_{\wb})_{\ell\pr\ell} \eqdef
-i\big(\psi_\ell^{u}, \,\partial_{\wb}\,\psi^u_{\pr\ell}\big)
 = -i \big(\psi_\ell^{u},\psi^u_{\pr\ell}\big) (h^u)^{-1}\partial_{\wb}\, h^u\,
= {i\pi\over 2k}w \,\dl_{\ell\pr \ell}^{{\rm mod}k}}
is a reducible $U(k)$ gauge field with constant field
strength on the dual torus. The dual torus is given by
\eqn\dualtorus{
\hat{\hbox{\af T}}\,^2=\hbox{\R}^2/\,\hat{\Lambda},}
where with our choice for the shift in \nnabel\
the dual lattice $\hat{\Lambda}$ is generated
by the two orthonormal vectors $\hat{e}_\mu=e_\mu$.
In real notation the potential $\hat A$ takes the simple form
$${\hat A}_1 = {\hat A}_w + {\hat A}_{\wb} =
-{{\pi}\over k}u^2\hbox{\id}\quad,\quad
{\hat A}_2 = i({\hat A}_{\wb} - {\hat A}_w) = {\pi\over
k}u^1\hbox{\id}.
$$

The transformed gauge potential is only apparently Abelian. The
non-Abelian character of this $U(k)$ bundle can be seen from the
peculiar boundary conditions of the holomorphic structures
induced by ${\hat A}$. The corresponding transition functions
which relate $u$ and $u+e_\mu$ on the dual torus, \eqn\bcnnn{
\psi^{u+e_1}(x)=\hat U_1 \psi^u(x)\quad,\quad
\psi^{u+e_2}(x)=\hat U_2 \psi^u(x)} are determined by the shift
identity \trans\ and the transformation properties \zerotransu\
as follows, \eqn\nahmtranss{ \hat U_1=\exp{2\pi ix^1+{i\over
k}\pi u^2}\hat S_1\mtxt{and} \hat U_2=\exp{2\pi ix^2-{i\over
k}\pi u^1} \hat S_2,} where $$ \hat S_1=S_2^{-1}\mtxt{and} \hat
S_2=S_1.$$
Recall that the non-Abelian elements $\hat S_\mu$ generate a
finite non-Abelian  subgroup of $U(k)$
\eqn\invol{\hat S_\mu^k=\hbox{\id},\qquad \hat S_1\hat S_2
=\exp{-i\,2 \pi/k}\; \hat S_2\hat S_1.}
The last relation guarantees that for any fixed $x$
the $\bfpsi^u$ are sections of a $U(k)$-bundle over the
dual torus $\hat{\hbox{\af T}}\,^2$ with coordinates $u$:
\eqn\cocycle{ \hat U_2(u\!+\!e_1)\,\hat U_1(u)
= \hat U_1(u\!+\!e_2)\, \hat U_2(u).}

The first Chern class of this bundle follows from the fact that
the (non-Abelian) Nahm transformed gauge potential ${\hat A}$ is
just $k$ times any of its diagonal elements, hence
$$\int_{\hbox{\afs T}^2}{\rm tr}\, F(\hat{A}^u)= 2\pi\Longrightarrow
c_1 (\hat{A}^u)=\hat k=1.$$
It can be shown that the Nahm transform of $\hat{A}$ is $A$, i.e.
the Nahm transformation is involutive. It is a particular case of
the more  general Mukai transform defined for
holomorphic sheaves (which do not necessarily define holomorphic
bundle structures)\mmuk\ref\muk{S.~Mukai, Adv.~Stud.~Pure Math. {\bf
10} (1987) 515.}\foot{see also \ref\hbr{U. Bruzzo and F. Pioli,
Diff. Geom. Appl., {\bf 14} (2001) 151-156.} for a view closer to
physical applications}.

\newsec{Instantons in {\af T}$^2$}

Let us consider an instanton field $\Psi$ on the torus with
charge $k$, that is,  a solution of \covholomorphy\ subject to the
boundary conditions \bc\ . The associated $U(1)$ gauge potential
$A$ is a connection defined in a line bundle $E_k$ with first
Chern class $c_1(E_k)=k$. The holomorphic structure associated to
$A$ in $E_k$ is {\it globally} \  equivalent to one of the $A^u$
described in the previous section. This means that there is a
global (periodic) complex gauge transformation $g$: {\af
T}$^{\,2}\to$ \C$\,\setminus \{0\}$ such that
\eqn\gpot{ A_\zb= g\,\big(A^u_\zb +i\,\pa_\zb\big)g^{-1}\mtxt{and}
D_\zb=g\,(\pa_\zb-i A^u_\zb)\,g^{-1}.}
Therefore, up to U(1) gauge transformations the $N\!+\!1$
components of $\Psi$ can be expressed in terms of the $k$
independent solutions $\psi^u_\ell$ of the zero mode equation
\szm\ as follows,
\eqn\inst{
\Psi={1\over \sqrt{\bfpsi^{u \dagger} \bA^\dagger \bA \bfpsi^u}}\
\bA\bfpsi^u,\mtxt{where}
\bA  =\pmatrix{a_0^{\;\,1}&a_0^{\;\,2}&\cdots&
a_0^{\;\,k}\cr a_1^{\;\,1}  & a_1^{\;\,2}  &\cdots & a_1^{\;\,k}
\cr \cdot &\cdot&\cdots&\cdot\cr \cdot &\cdot&\cdots&\cdot\cr
a_N^{\;\,1}  & a_N^{\;\,2}&\cdots & a_N^{\;\,k}  \cr }.}
Hence, any instanton solution is characterized by a point $u$
in the dual torus $\hat{\hbox{\af T}}\,^2$ and
a $(N\!+\!1)\!\times\! k$ matrix
$\bA$
subject to certain constraints given below.
This characterization provides a constructive method
to describe the moduli space of instantons with charge $k$.

The projective nature of the sigma fields $\Psi$ implies that
matrices  differing by a non-vanishing multiplicative factor must
be considered as equivalent,
\eqn\equivalent{ \bA\sim \lambda \bA,\qquad
\lambda\neq 0,}
because they give rise to the same instanton
field. Furthermore, in order to satisfy the sigma model condition
$\Psi^\dagger(x) \Psi(x)\!=\!1$ the $\bA\bfpsi^u$ should never
vanish ($\bfpsi^u\in $\C$^k$ never vanishes since the
$\psi^u_\ell$ have no common zeros) and this imposes a constraint
on $\bA$. Finally, because of the boundary conditions \bcnnn\ we
have the identifications
%the parameters $(u,\bA)$ and
%
\eqn\ident{(u,\bA)\sim \hat{T}_\mu(u,\bA)= \big(u\!+\!e_\mu\,,\,\bA
\hat U_\mu^{-1}\big),}
since the two pairs give rise to the same $\Psi\in\;$\cpn and
hence must be identified. There is no further identification
since a shift $u\to u+\al$ with $\al\notin\,$\Z$^2$ cannot be
compensated by a (necessarily) unitary matrix. This would not be
compatible with ${\bfpsi}^u$ being a section of the $U(k)$-bundle over
the dual torus with charge $1$.

In order to understand the remaining constraint on the ${\bA}$
matrices let us consider a simple example.
It is the {\it basic instanton} of charge $k$ of the \cp$^{k-1}$
model defined by the basis \szm\
of zero-mode sections of $E_k$,
\eqn\triv{\Psi_\ast^u={1\over
\sqrt{\bfpsi^{u\dagger}\bfpsi^u}}\;\bfpsi^u.}
In the $(\bA,u)$ parametrization
this solution corresponds to $\Psi_\ast^u=(u,\hbox{\id}_k)$.
Notice that in this case the constraint is satisfied because
sections of the basis \szm\ do not have a common zero
\ref\aae{M.~Aguado, M.~Asorey, J.~G.~Esteve, Commun. Math. Phys.
{\bf 218} (2001) 233}.

It is not hard to find sufficient conditions on $\bA$ for $\Psi$
to be normalizable. Clearly, the denominator in \inst\
$$  \big(\bfpsi^{u\dagger}\,\bA^\dagger \bA \bfpsi^u\big)^{1/2}
$$
is never zero if det$(\bA^\dagger \bA)\neq 0$. Since the rank of
the $k\times k$ matrix $\bA^\dagger \bA$ is less or equal that
min($k$, $N\!+\!1$), this can only be fulfilled for $k\leq
N\!+\!1$. Hence in this case the maximal rank condition is {\it
sufficient}, i.e.
\eqn\sufcond{ \ker\left\{\bA\,:\,\hbox{\C}^k \to
\hbox{\C}^{N+1}\right\}=0 \qquad {\rm if\ }k\leq N+1.}
However, even in that case this condition is not necessary. The
fact that $\Psi(x)$ has to be a non-null vector for any point $x$
requires that the matrix $\bA$  be viewed as a projective map
{\C}P$^{k-1}\to\;$\cpn \ from rays of {\C}P$^{k-1}$ into rays of
\cpn . This just means that the kernel of $\bA$ must not lie in
the image of $\bfpsi^u(x)$ for any $x$ on the torus, that is
\eqn\constr{{\rm range}(\bfpsi^u)\cap \ker(\bA)=\emptyset.}
Otherwise  $\bA \bfpsi^u$ will have zero norm at some point and
will not be a true sigma model field. Notice the compatibility of
this constraint with the identifications \ident. This concludes
the characterization of instanton solution and provides an
explicit procedure for a global description of the moduli space.

Before discussing the subtleties related to \constr\ in the
general case we consider some simple examples of moduli
spaces $\CM^N_k$. First of all is clear that
$$\CM_0^N=\hbox{\cpn}\qquad {\rm and} \qquad \CM_1^N=\emptyset
\mtxt{\rm for} N>0. $$
In the first case because $\Psi^u=0$ for $u\neq 0$ and $\Psi^0$
is an arbitrary constant vector in \cpn. The second case follows
from the fact that for $k=1$ there is only one zero mode $\psi^u$
which has exactly one zero on {\af T}$^2$. Then all $N\!+\!1$
components of $\Psi$ would vanish at this point and hence it
could not be normalized.

A simple non-trivial case where the moduli space can completely
be constructed is $\CM_2^1$, that is, the charge $2$ sector of the
{\af C}P$^1$ model. Since the range of the basic instanton
$\Psi_\ast^u$ completely covers {\af C}P$^1$ the constraint
$\constr$ if fulfilled if and only if the matrix $\bA$ is
regular. In this case the sufficient condition \sufcond\ is also
a necessary one. Since  $\bA$ maps into \cp$^1$ we may impose
det\ $\bA\!=\!1$ and identify $\bA$ with $-\bA$. It follows that
the equivalence classes of matrices are to be regarded as elements
of SL(2,\C)/\Z$_2=\,$PSL(2,\C).

Because of the identifications \ident\ the moduli $\CM_2^1$ is
just a non-trivial bundle over the dual torus (with coordinates
$u$) with fiber PSL(2,\C). The bundle structure is determined by
the coset defined by the lift  of the action of the discrete
translation group \Z$\times$\Z\ on the bundle
$\hat{\hbox{\C}}\times$ PSL(2,\C), given by %\ident
\sp:
\eqn\twt{\hbox{\Z}\times\hbox{\Z}=\{(\hat T_1^{\,n_1},\hat
T_2^{\,n_2});\ n_1,n_2\in \hbox{\Z}\},}
where $\hat T_1$ and $\hat T_2$ are the basic generators defined
in \ident.
The final result is that
\eqn\modab{\CM_2^1={\hat{\hbox{\C}}\times
\hbox{PSL(2,\C)}\over \hbox{\Z} \ \times\  \hbox{\Z}}. }
In the general case the construction of the moduli space is more
involved since the solutions of the constraint \constr\ are not so
explicit. But once we have identified the embedding of the
space-time torus {\af T}$^{\,2}$ into {\af C}P$^{k-1}$ given by the
basic instanton $\Psi_\ast^u$, the set of allowed matrices can be
parametrized as follows:
The basic instanton solution $\Psi_\ast^u$ defines a map
\ {\af T}$^{\,2}$ $\to\,$\C$^k$. Consider all linear subspaces
$V_n$ of \C$^k$ of dimension $n\!<\!k$ having empty intersection
with range$(\Psi_\ast^u)$. The space of matrices $\bA$ which
define regular instantons for a fixed $u$ can be identified with
the pairs $(V_n,\hbox{\bf B})$ defined by the subspaces $V_n$ and
the non-degenerate linear maps {\bf B} mapping the orthogonal
complement $V_n^\perp$ of $V_n$ into the target space \C$^{N+1}$.

This means that the moduli space of instantons can be identified
with a bundle over the dual torus with fiber isomorphic to the
product $\CV\times\hbox{PL}_0(k\!-\!n,N\!+\!1)$  of the set $\CV$
of $V_n$ subspaces and the set PL$_0(k\!-\!n,N\!+\!1)$ of
non-degenerate projective maps from {\C}P$^{k-n-1}$ into
\cp$^{N}$. The bundle is defined by modding out the trivial bundle
$\CV\times\hbox{PL}_0(k\!-\!n,N\!+\!1)\times\hat{\hbox{\af C}}$
by the lift of the discrete translation group \Z$\times$\Z\ given
by \ident.

To illustrate how the construction works let us consider a simple
non-trivial case in some detail: $\CM_3^2$. In this case we have a
dense subset $\CM^2_{3(0)}$ which is given by the bundle over the
dual torus with fiber PSL(3,\C)=SL(3,\C)/\Z$_3$, the equivalence
classes of 3$\times$3 matrices with det$\ \bA=1$. The complex
dimension of the sub-manifold, dim $\CM^2_{3(0)}=9$, equals  that
of the total space ${\CM_3^2}$. However there is another
sub-bundle in ${\CM_3^2}$ with lower dimension. The fibers of
this sub-bundle $\CM^2_{3(1)}$ are the classes of 3$\times$3
matrices with one-dimensional kernel $V_1$ which does not
intersect the image of the map $\bfpsi_\ast^u:\hbox{\af
T}^2\to$\C$^{\,3}$. The complex dimension of this subbundle is
six. The total space is the union of the two strata,
\eqn\stratt{
{\CM_3^2}=\CM^2_{3(0)}\cup \CM^2_{3(1)}.}
the second being the boundary of the first one. Notice that the
subset of the second stratum  $\CM^2_{3(1)}$ associated to a fixed
kernel can be identified with $\CM_3^1.$

\newsec{Global properties of the moduli of instantons}
The complex dimensions of the moduli spaces are
$$\dim \CM_k^N=(N\!+\!1)\,k $$
as follows at once from our matrix representation
of the \cpn-fields. Note that this number is
invariant under the interchange of the instanton number $k$
and the number $N\!+\!1$ of sigma field components.

In the case $k\geq N\!+\!1$  there is a natural stratification
\nref\ccr{R.~Catenacci, M.~Cornalba, C.~Reina,
Commun.~Math.~Phys.~{\bf 89} (1983) 375.} of the moduli
spaces,
\eqn\str{ \CM_k^N=\bigcup_{n=k-N-1}^{k-2}\;\;\CM^N_{k(n)}\;,}
according to the dimension $n=\dim \ker {\bf A}$, but this does
not mean that the moduli space is not a smooth manifold. From the
matrix parametrization it is obvious that $\CM_k^N$ is smooth,
although it might seem hidden by the stratification \str\
introduced in Ref. \ccr. Moreover, $\CM_k^N$ is  a K\"ahler
manifold \Ruback\ and the associated Riemannian structure plays an
important role in accurate semiclassical expansions of \cpn\
models \sp.  The matrix parametrization permits to analyze the
global features of these geometric structures, in particular, the
incompleteness of the Riemannian metric, but we shall only focus
into the analysis of the global topological structure of these moduli
spaces. Below we summarize some of the relevant results.

 $\CM_k^N$ is
non-empty and connected  for any $k> 1$ and $N > 1$, 
i.e.  $\pi_0(\CM_k^N)=0$. 
$\CM_0^N=${\C}P$^N$, and $\CM_k^1=\emptyset$ for $k> 1$.
The simplest non-trivial moduli space is $\CM_2^1$, and because of
the identification \modab\ we have
\eqn\upi{\pi_1(\CM_2^1)=\hbox{\Z}_2\ \times\ \hbox{\Z} \ \times\
\hbox{\Z}.} The next case $\CM_2^N$ has also a non-trivial bundle
structure over the dual torus. Its fiber $E_u$ is the
projective set PL$(2,N+1)$ of equivalence classes $(N\!+\!1)\times
2$ matrices $\bA$ with det$\bA \bA^\dagger\neq 0$, which can be
identified with the subset of the projective space
{\C}P$^{2(N+1)-1}$ defined by excluding rays of the form
$(\lambda_1\vec{a}, \lambda_2\, \vec{a})\in
\hbox{\C}^{N+1}\times\hbox{\C}^{N+1}$. This fiber $E_u=$PL$(2,N)$
is homeomorphic to {\C}P$^N\times\hbox{\C}^{N+1}$. The bundle
structure is defined by the coset defined by the lift of the
action of the translation group \Z$\times$\Z \ to the bundle
$\hat{\hbox{\C}}\times $PL$(2,N)$ given by \twt\ . Thus,
\eqn\dpi{\pi_1(\CM_2^N)= \hbox{\Z}_2\ \times\ \hbox{\Z} \ \times\
\hbox{\Z}.}
A more complex moduli space is $\CM^1_{k}$. In this case the
fiber $E_u$ is identified with the (projective) set PL$(k,2)$ of
$2\!\times\!k$ matrices $\bA$ with maximal rank 2 whose kernel
has an empty intersection with $\Psi^u_\ast(x)$ for any $x$ on
the torus. In  terms of the nodes of the corresponding \cp$^1$
field, the fiber $E_u$ can be seen as homeomorphic to the set of
$k$ pairs $(a_i,b_i)\in \hbox{\af T}^2\times\hbox{\af T}^2$ (for
$i=1,2,\cdots, k$), satisfying the following  constraints
\eqn\conssum{ \sum_{i=1}^{k} a_i\in \big\langle u+{\ft{k}{2}}
e\big\rangle,  \sum_{i=1}^{k} b_i\in \big\langle
u+\ft{k}{2}e\big\rangle \mtxt{and} a_i\neq b_j\;\;\; \forall
i,j.}
 That means that the sum of the zeros are the same (modulo
$\Lambda$) for both components of the \cp$^1$ field since they
fulfill identical boundary conditions. The space of nodes of any
of the two components of $\Psi$ is homeomorphic to
\cpt$^{k-1}$ (see appendix).  The space $\CZ$ of zeros of the two
components of $\Psi$ satisfying the constraint of having no
common nodes  is a bundle over
$\hbox{\cp}^{k-1}$ with fiber $\hbox{\cp}^{k-1}-{\cal C}$, where
$$ {\cal C}=\big\{\{a_i,b_i\}\in \hbox{\af T}^2\times \hbox{\af
T}^2\big\vert \,i=1,2,\dots,k, \mtxt{with} a_i= b_j\mtxt{for
some} i,j\big\}.$$ The first homotopy group of $\CZ$ is
 $$ \pi_1(\CZ)=\hbox{\af F}_{2}\times\hbox{\af
Z}^{k-3}, $$
where \F$_2$ is the free group with $2$ generators, i.e. the
first homotopy group  of a bouquet of two circles $S^1\vee S^1$.
This follows from the following characterization of ${\cal C}$:
The second component $\Psi_2$ of the sigma field $\Psi$ must be a
holomorphic section of the line bundle $E_k$ without common zeros
with the first component $\Psi_1$. Let $a_1, a_2,\dots,a_k$ be
the  zeros of $\Psi_1$. There is only one constraint  \conssum\
on the position of these zeros. It is always possible to choose a
different point $b$ in \T$^2$  such that $b\neq a_i$ for any
$i=1,2,\dots,k.$ It is obvious that the points
$a_1,\dots,a_{k-1},b$ will never satisfy the constraint \conssum.
Then, the space of vectors ${\hat{\cal C}}$ whose rays are in
${\cal C}$ is given by all vectors in the subspaces $\Psi(a_1)=0,
\Psi(a_2)=0,\dots ,\Psi(a_{k})=0$. The space of  all holomorphic
sections of $E_k$ is parametrized by the coordinates $(\alpha_1,
\alpha_2\cdots\alpha_k)$ defined by $\alpha_1\!=\!\Psi(a_1),\dots
,\alpha_{k-1}\!=\!\Psi(a_{k-1}),\,\alpha_k\!=\!\Psi(b)$. In this
parametrization ${\hat{\cal C}}$ is made out of the first $k-1$
coordinate hyperplanes $\alpha_i=0\,,i=1,\dots,k\!-\!1$ and the
extra hyperplane $\Psi(a_k)=0$. Then, \cp$^{k-1}-{\cal C}$ can be
identified with \C$\times$\C$_\ast^{k-3}\times$\C$_{\ast\ast}$, where
\C$_{\ast\ast}$ denotes the complex plane \C\ without two points
$0,1$. From this construction it is obvious that
$\pi_1(\hbox{\cp}^{k-1}-{\cal C})= \hbox{\af F}_{2}\times\hbox{\af
Z}^{k-3}$. Then, the first homotopy group of constrained $2\times
k$ matrices PL$(k,2)_c$ has a non-trivial non-Abelian homotopy
group $\pi_1($PL$(k,2)_c)=\hbox{\F}_{2}\times\hbox{\Z}^{k-3}\times
\hbox{\Z}_2$. This implies that the first homotopy group of the
moduli space $\CM_{k}^1$ is
\eqn\dppi{\pi_1(\CM_{k}^1)=\hbox{\F}_{2}\times \hbox{\Z}^{k-1}
\times\hbox{\Z}_2.}
These topological properties of moduli spaces of instantons \dpi\
and \dppi\ are very different which will be in contradiction with
the existence of any kind of  Nahm transform for \cpn sigma
models. This turns out to be the major physical consequence of the
results of this section.

The fact that the space of unit charge instantons is empty
is reminiscent of a similar property of the orbit space of
Yang-Mills fields with charge one instantons on the torus {\af
T}$^4$ \mmuk\BvB. In that case, it appears as a consequence of the
existence of Nahm transform and the fact that there are no
$U(1)$ instantons.
This analogy suggests that perhaps the same property for \cpn\
instantons can be derived from a similar duality transform.
In addition, the dimensions of $\CM_k^N$ and $\CM_{N+1}^{k-1}$
are the same.

A first indication that the Nahm transform might not exist for
the \cpn\ models arises from the fact that the transform ${\hat
A}_z$ of the Abelian potential $A_z$ associated with $\Psi$ is
non-Abelian and thus cannot be associated to a \cpn\ field on the
dual torus. One direct way of checking whether such a duality
exists is to compare topological properties of $\CM_k^N$ and
$\CM_{N+1}^{k-1}$.

Now, the topological structures of
$\CM^N_2$  and  $\CM^1_{N+1}$ given by
\dpi\ and \dppi\ are very different for $N\!>\!1$. %\sss\
This already allows us to exclude the existence of an
invertible Nahm transform, at least in these cases.
The same topological non-equivalence holds for more general moduli
spaces, which excludes the existence of a generic duality
transformation. The only case where these topological arguments
fail to exclude the existence of a (generalized) Nahm transform
is the selfdual moduli spaces $\CM^N_{N+1}$ because of the trivial
identity between both moduli spaces.

\overfullrule=0pt

\newsec{Compactification of moduli spaces and singular instantons.}

The moduli spaces of instantons analyzed in the previous sections
have natural compactifications obtained by adding the boundaries
consisting of the matrices $\bA$ which do not satisfy the
constraint \constr. These configurations correspond to fields
which do have common zeros in all its components. Properly
speaking these are not \cpn\ fields because at these common
points they do not define maps {\af T}$^2\to\,$\cpn. These points
can also be seen as  singular points when one introduces the
normalization factor to have a unit norm representation of the
field $\Psi$. These singular points can be interpreted as centers
of singular instantons.
It is envisable to consider %the violation of the constraint by
the existence of common zeros as an effective charge reduction
induced by the appearance of singular instantons. This
observation provides additional information about the structure
of the boundary of ${\CM}^N_k$ in ${\overline{\CM}}{}_k^N$.

The resulting moduli space $\overline{\CM}{}^N_k$ is compact and has a
bundle structure over the dual torus with compact fiber
\cp$^{k(N+1)-1}$. This compactified moduli space is stratified
according to the number of common zeros of the different
components of the field \lns. The generic dense stratum contains
all regular instantons. The other strata consist of  singular
instantons. The degree of singularity is parametrized by the
number of common zeros. This is reminiscent of a similar
phenomenon occurring in  Yang-Mills theory \pm.

\def\thuno{{\vartheta_1(z-z_0|i)}}
\def\bthuno{\overline{\vartheta_1(z-z_0|i)}}
\def\tps{\tilde\Psi}
A single singular instanton can be viewed as a regular one with
one topological charge less and with a pointwise singularity at a
point $x_0$. In fact, we can generate singular instantons by
adding such singularities to all regular instantons of lower
charges, which gives a complete characterization of the subspace
of singular instantons. Let $\Psi$ be  a (pointwise normalized)
\C$ P^N$ instanton configuration of charge $k.$ One can obtain a
singular instanton $\tps$ of charge $k\!+\!1$ by simply
multiplying each component of $\Psi$ by the phase of a theta
function carrying unit charge:
\eqn\newsingular{
\tps = {\thuno\over{|\thuno|}} \Psi. }
Observe that this phase has a singular point at the zero of the
theta function, $z=z_0.$ The corresponding holomorphic line
bundle structure is shifted to $u+z_0$.

The associated gauge potential splits into the previous, charge
$k$ piece and a new contribution coming from the singular phase:
\eqn\newsingamu{
{\tilde A}_{\zb} = - i\tps^\dagger \partial_\zb \tps = A_\zb + a_\zb,}
where
\eqn\newsinglittleamu{ a_{\bar z} = -i {\rm e}^{-i\,{\rm
arg}\,\thuno} \partial_{\bar z}\, {\rm e}^{i\,{\rm arg}\,\thuno}
= {i\over2}\,{{\partial_{\bar z}{{\bthuno}}}
\over{{\bthuno}}}. }
The additional  contribution to the topological charge density is
singular
\eqn\newsingtopoldens{
 {1\over{4\pi}} \varepsilon_{\mu\nu} f_{\mu\nu}(x)
= -\,{{2i}\over\pi} \partial_z a_{\bar z}(x) = {1\over{4\pi}}
\nabla^2 \ln {{\bthuno}} = \delta^{(2)}(x-x_0), }
corresponding to a unit point charge at $x_0$. \
Thus the total topological density is
\eqn\nsing{
 {1\over{4\pi}} \varepsilon_{\mu\nu} \tilde{F}_{\mu\nu}(x)
= {1\over{4\pi}} \varepsilon_{\mu\nu}
F_{\mu\nu}(x)+\delta^{(2)}(x-x_0).}
The same singular behavior appears in the
new distribution of the energy density and this
nicely illustrates the effect of including singular instantons.
%%%

Although the physical interest of singular configurations is not
yet understood, field configurations in the vicinity of singular
instantons appear in the regular space. To further clarify the
structure near singular instantons we shall consider some
interesting cases.

A regular instanton in the bulk of $\CM_3^1$ can be built by
choosing the parameters of the matrix $\bA$ in such a way that
the two components of the \cp$^1$ field have well separated nodal
points (e.g. the instanton configuration shown in fig.1.)

\ninepoint
 $\,$
{\hskip 1.5cm \epsfxsize=9cm\epsfbox{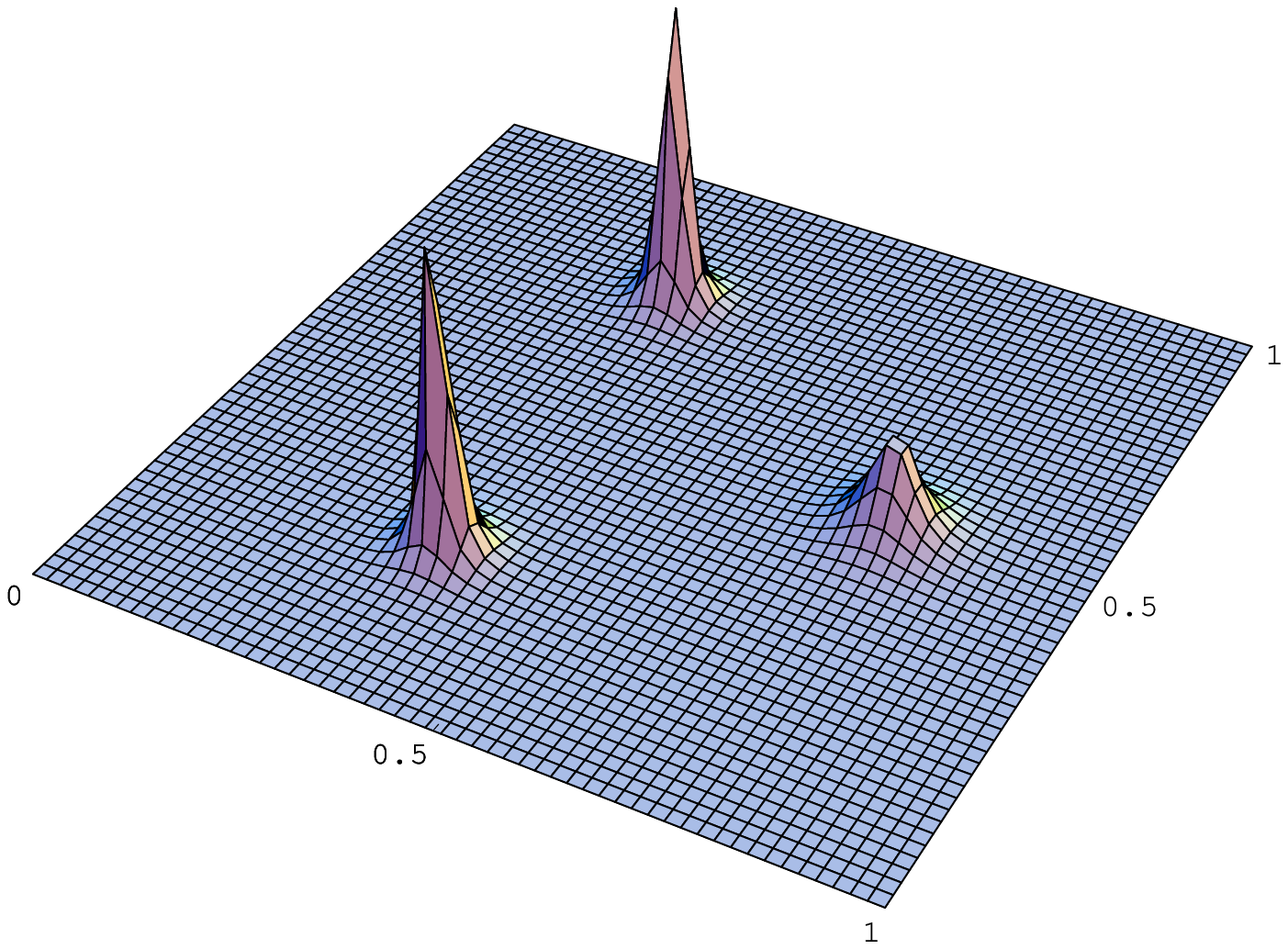}} \vskip 0cm
\vbox{\noindent{\bf Figure1.} Energy density distribution of a
{\afw C}P$^1$ regular instanton $u=0,A=\pmatrix{2\sqrt{2}&2&0\cr
0.9&0&0}$ with charge $k=3$  in $\CM_3^1$}
 \tenpoint

 We can approach the boundary of the moduli of regular instantons
by  choosing  parameters of $\bA$ in such a way that the nodal
points of the two components of $\Psi$ are very close to each
other. We are near a singular instanton and find peaks in the
energy density located on the nodal points showing the strong
localization of energy and topological charge on singular
instantons.

 $\,$
 \vskip -1.1cm
%$\,$
\ninepoint{\hskip 1.5cm \epsfxsize=9cm\epsfbox{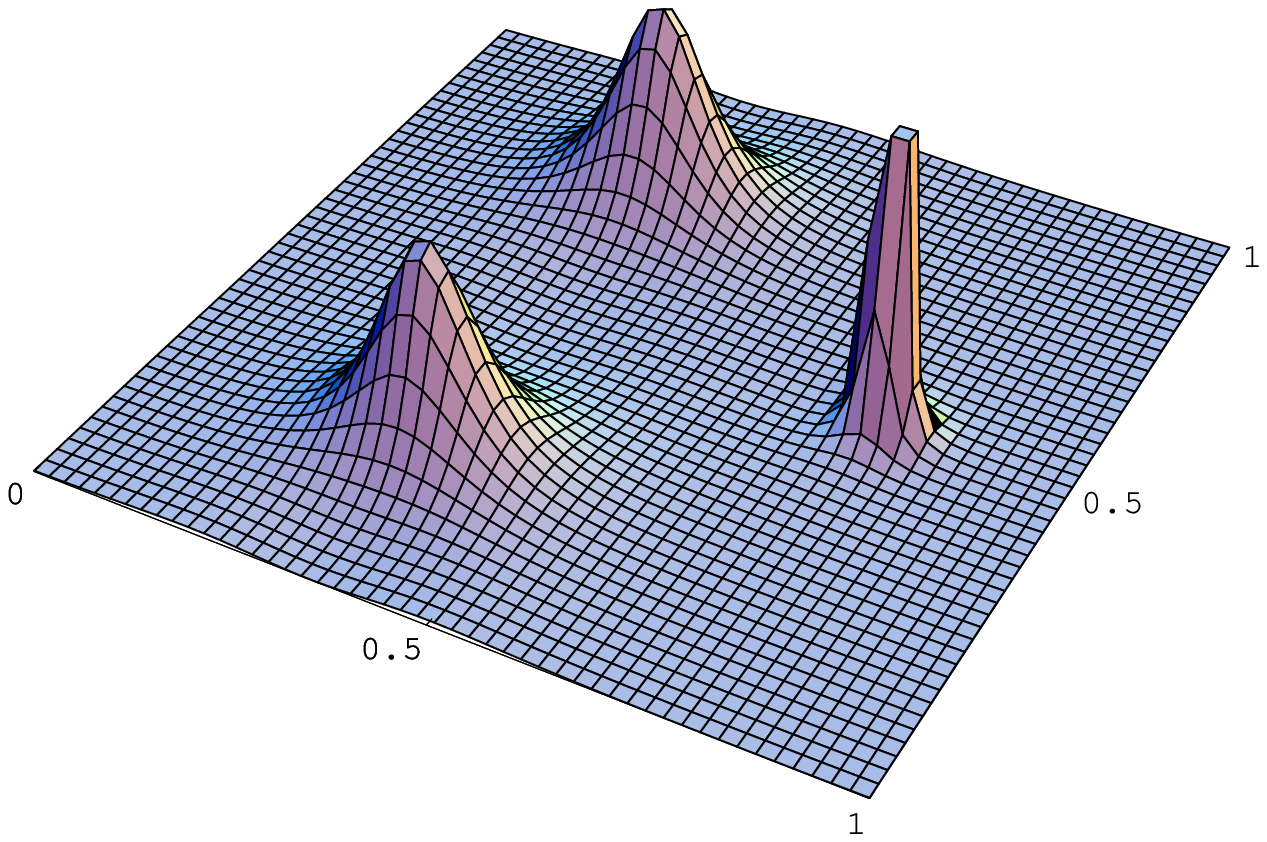}} \vskip
0.2cm {\vbox{\noindent{\bf Figure 2.} Energy density distribution
of a {\afw C}P$^1$ regular instanton $u=0,$
$A=\pmatrix{2\sqrt{2}&2&0\cr 0&0.1&-i}$ with charge $k=3$ close to
one singular instanton in $\CM_3^1$.}}
\tenpoint

In the limit case we obtain a charge 3 instanton   with   one
singular instanton and two regular ones. The picture is quite 
similar to Fig.2 with two lumps in the topological density, 
corresponding to two interacting instantons of charge 2 and 
finite size, and one singular instanton which is not shown in 
the numerical simulation.

Another way of approaching the boundary of the moduli space is by
choosing one of the components of $\Psi$ very small. In this case
we approach a completely singular instanton with $k$
singularities and one null component.

Now, the identification of single instantons in a multi-instanton
configuration is not always clear. In fact, there are strongly
overlapping configurations where it is hard to identify the
constituent instantons. In Fig. 3 the configuration seems to
contain four lumps whereas its total charge is $k=2$.

\ninepoint
 $\,$
 \vskip -1.4cm
% \vskip 9.7cm
{\hskip 1.5cm \epsfxsize=9cm\epsfbox{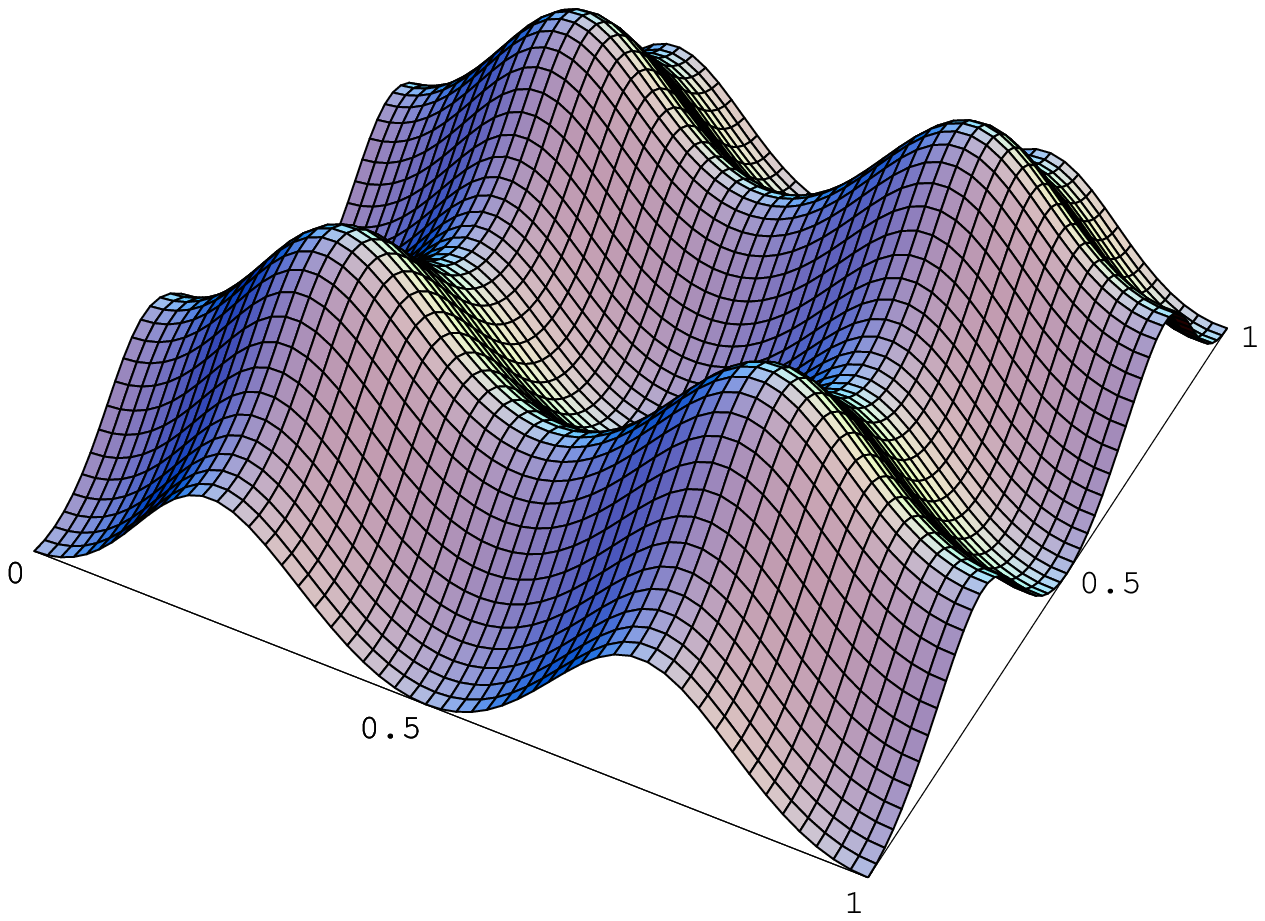}} \vskip 0.0cm
{\vbox{\noindent{\bf Figure 3.} Energy density distribution of a
{\afw C}P$^1$ regular instanton $u=0,A=\pmatrix{1&0\cr 0&1}$ with charge
$k=2$ in $\CM_2^1$}} \vskip 0cm

 \tenpoint
 Moreover, the interaction between instantons can be very involved
and we can find densities with the shape of a volcano as in Fig. 4
which seems to describe a ring of instantons whereas its total
topological charge is $k=2$.

 \vskip -1.2cm
\ninepoint
 {\hskip 1.5cm \epsfxsize=9cm\epsfbox{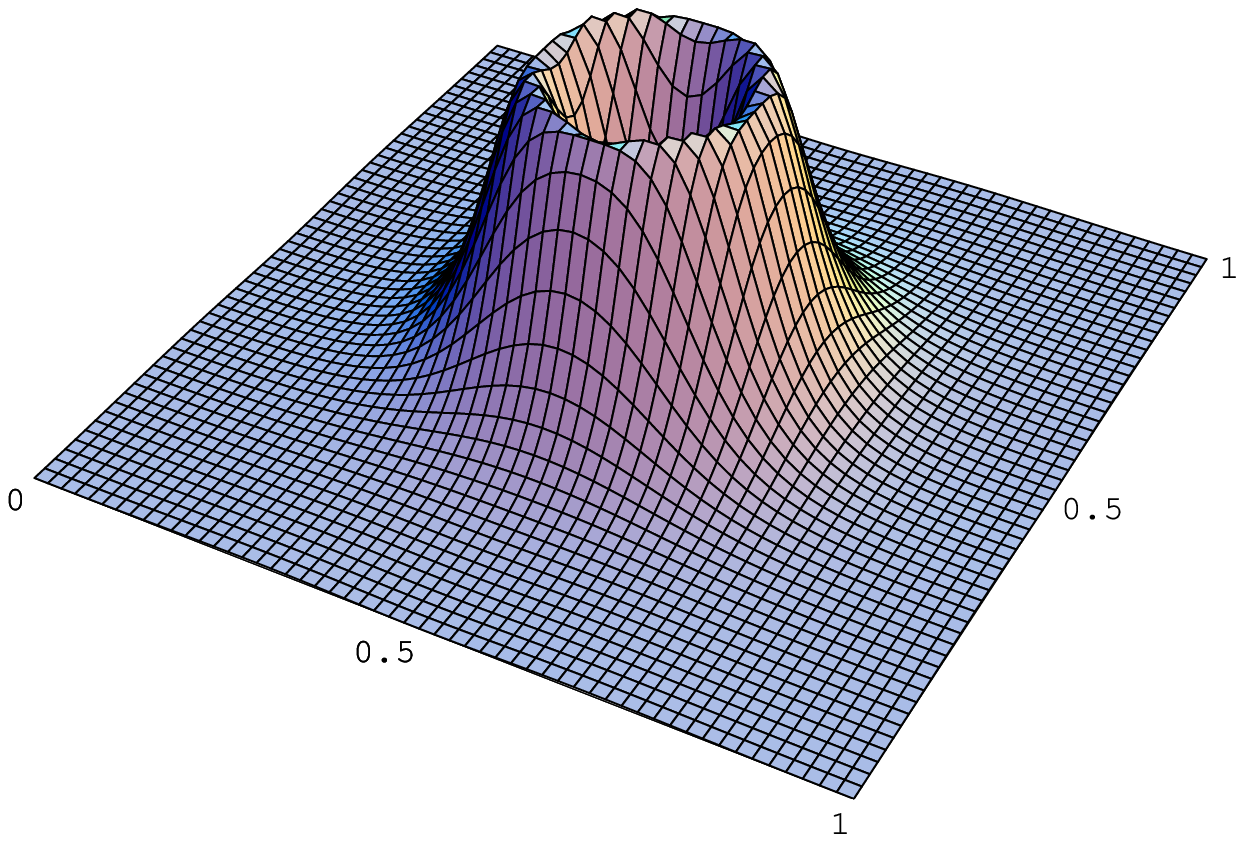}}
\vskip -.0cm {\vbox{\noindent{\bf Figure 4.} Energy density
distribution of a {\afw C}P$^1$ strongly interacting regular instanton
$u=0,A=\pmatrix{0.42-0.72i&0.27-0.47 i\cr -0.49-0.85i &-0.10-0.18i
}$ with charge $k=2$ in $\CM_2^1$}}
\vskip 0.5cm
\tenpoint

Strongly interacting instantons dominate in dense gas regimes
whereas isolated instantons are more relevant for dilute gas
phases.  In general it is very dificult to identify the number of
instatons of a single configuration in dense regimes. This
difficulty  increases in numerical simulations where the leading
configurations are not exact selfdual solutions and  are made of
instantons and anti-instantons.

 In any case, singular instantons appear as
a limiting case of small size instantons. In particular, they are
responsible for  the geodesic incompleteness of the moduli space
of regular instantons \sp. Such configurations become of physical
interest in supersymmetric \cpn\ theories where they are
important to localize topological Green functions in the
corresponding topological field theory \lns.

We close with some more general remarks about the possible role
of singular instantons: The semiclassical expansion in a dilute
gas of instantons has a different behavior depending on the
number of components of the sigma model. For $N\!>\!1$ the
expansion is dominated by large instantons and we cannot trust
this approximation to describe the deep infrared behavior of
the theory \ref\lush{M.~L\"uscher, Nucl.~Phys.~{\bf B200} (1982)
61.}. In particular, their relevance for the confinement
mechanism is unclear. On the other hand, for the \cp$^1$ model
the expansion is dominated by instantons of small sizes because
of the different $\beta$ function \lush. In the extreme case,
singular instantons carry the leading effects and this is
self-consistent with the dilute gas approximation. However, an
ultraviolet regularization is in any case necessary. In lattice
regularization the size of  small instantons is bounded below by
the lattice spacing and as a consequence the scaling properties
of the topological susceptibility are changed. This leads to
difficulties when one approaches the continuum limit
\lush\ref\bur{M.~Blatter, R.~Burkhalter, P.~Hasenfratz,
F.~Niedermayer, Phys.~Rev.~{\bf D53} (1996) 923.}. The above
discussion indicates that a continuum approach is feasible.
Although similar effects are expected to occur, they might be less
severe and lead to the stabilization of the ultraviolet
catastrophe seen in the lattice approach.

In some sense the appearance of singular instantons is a dual
effect of the existence of reducible instantons. A \cpn\
instanton is reducible when it can be considered as living in a
lower dimensional \cp$^{N-1}$ projective subspace of \cpn.
Reducible instantons belong to the strata of $\CM^N_k$ associated
with matrices $\bA$ with rank lower than $N$. In the compactified
moduli space $\overline{\CM}{}^N_k$ there are two classes of strata, one
corresponding to reducible instantons and the other to singular
instantons. In one case there is a charge reduction and in the
other a dimension reduction. The role of the two kinds of strata
are interchanged when we compare the dual cases $\CM^N_k$ and
$\CM^{k-1}_{N+1}$. If we exclude both types of instantons we are
left with the modular space of generic regular instantons. The
global structure of the space of generic regular instantons in
$\CM^{k-1}_{k}$ is much simpler. It is always a bundle with the
dual torus as basis and as typical fiber the group PSL$(k,$\C),
twisted by the boundary conditions \ident.

\newsec{Conclusions}
The global structure of the moduli space of instantons in the
\cpn\ model on a torus has a more explicit description than for
Yang-Mills theory.  However, this by no means implies that its
geometrical and topological properties are simpler. In fact, in
the case of gauge theories there exists a Nahm transform
establishing a one-to-one correspondence between two a priori
very different moduli spaces of instantons, $\CM^N_k$  and
$\CM^{k-1}_{N+1}$. We have shown that such a map cannot exist for
the \cpn\ sigma models.

We have identified the boundary of the space of regular
instantons with the space of singular instantons. This
identification of singular instantons as boundary configurations
of the space of regular instantons provides a new approach to the
analysis of the physical role of overlapping instantons in a
dense gas and in topological field theories. The role of
instantons in the confinement mechanism seems to be very
different in the  \cp$^1$ model and higher $N$ models. The
dominance of small or large instantons indicates a critical
transition or crossover between these two regimes. The behavior of
the theory in the presence of a $\theta$ term is also very much
dependent on the regime and size of leading instanton
contributions. In particular, the \cp$^1$ model shows a second
order phase transition at $\theta=\pi$ whereas in higher $N$
models it is not yet known whether a similar transition exists or
not. It is very plausible that the instantons will play a role in
the presence or absence of such a transition. If that case the
global structure of $\CM^N_k$ analyzed in this paper is expected
to have interesting physical effects.

\

\bigskip {\bf{Acknowledgements}}.\baselineskip16pt

We thank Ugo Bruzzo, Mario Escario, Antonio Gonz\'alez-Arroyo 
and Pierre van Baal 
for discussions. A. Wipf
thanks the University of Zaragoza and the MPI in Munich, where
part of the work has been done, for hospitality. M.  Aguado
is supported by a fellowship of MEC (Programa FPU). M. Aguado
and M. Asorey are partially supported by MCyT under grant
FPA2000-1252. This work was carried out in the framework of a
DAAD-MCyT grant. 
\listrefs

\appendix{$\!\!$}{Nodal structure of holomorphic sections}

Here we establish the connection between two different geometric
characterizations of the space of holomorphic sections of a
complex line bundle which have been used in the paper. First, it
is clear that this space  is  linear. On the other hand the
holomorphic sections are characterized up to a  constant by its
zeros (divisors). From the relation between the two approaches  it
follows that the  space of zeros of non-trivial holomorphic
sections has a projective space structure. Let us discuss in
detail how this projective structure emerges.

Indeed, the space of holomorphic sections of a complex line bundle
on the torus with Chern class $k$ is a linear space isomorphic to
\C$^k.$ Let us consider the basis introduced in \holobasis,
\eqn\appendixholoparts{ \chi_\ell^w(z)
=\chi_\ell(z-\ft{i}{k}w),\qquad \chi_\ell(z)=\exp{\pi k z^2/2}
\thetafnn{ z\!+\!{\ell\over k}} {0} {ik},}
for the holomorphic structure defined by $w=u^1\!+iu^2$ in
$E_k(${\af T}$^{\,2},$\C).
With this choice for the basis any holomorphic section in $E_k$ is
given by the expansion coefficients $c_\ell$ in
\eqn\appendixintermsoftheta{ \psi(z) = \sum_{\ell=1}^k c_\ell\,
\chi_\ell^w (z), }
The nodes of $\chi^w_\ell(z)$ are simple zeros and define the following
lattice
\eqn\appendixlatticeofzeros{
z_{m,n} = {i\over k}w - {\ell\over k} + \big( m+{1\over2} \big)
+ \big( n+{1\over2} \big){i\over k}, \qquad m,\,n\in\hbox{\Z}.}
Hence, there are $k$ such zeros in the fundamental domain and the
section belongs indeed to the bundle of charge $k.$
The boundary conditions
\eqn\appendixholobc{
\chi_\ell^w(z+1) = \exp{\pi k(z+1/2-iw/k)} \chi^w_\ell(z)\mtxt{and}
\chi^w_\ell(z+i)
= \exp{-i\pi k(z+i/2-iw/k)} \chi^w_\ell(z)}
have been derived in the main body of the paper.
They depend on $w\in\hat{\hbox{\af T}}\,^2$
which has been introduced to shift the gauge potential.

To discuss the topology of the \cpn\ fields it
is advantageous to use an alternative parametrization of the
sections for which the nodal structure is
explicit (but linearity is not).
It is given by the product representation
\eqn\appendixproducts{
\Theta(z) = \prod_{i=1}^k \exp{{\pi\over 2}\{(z-a_i)^2 +(1-i)z\}}
 \theta_1\big( z\!-\!a_i |\,i \big),\qquad
\vartheta_1(z|i)=\sum_n (-)^n\exp{-\pi(n+z-{1\over 2})^2}.}
Each $\vartheta_1(z-a_i|i)$ has zeros at the points of the lattice
$a_i+ \Lambda$, i.e. only a simple zero within the fundamental
domain. Coalescence of some of these zeros is allowed. $\Theta$
fulfills the boundary conditions
\eqn\appendix{
\Theta(z+1) = \exp{\pi k(z+1)\,-\,\pi a+i\pi k/2}
\;\Theta(z)\;\;,\;\;
\Theta(z+i) = \exp{-i\pi k(z+i)\,+\,i\pi a-i\pi k/2}\;\Theta(z),}
where $a=\sum a_i$.
They must coincide with those in \appendixholobc\ in order
to have a parametrization of the same space.
This gives rise to the following constraint on the
$a_i$ in \appendixproducts
\eqn\appendixconstraintonzeros{
\sum_{i=1}^k a_i = iw+\ft{k}{2}\,(1+i).}
It is easy to see that this sum is identical to that
obtained for the nodes in \appendixlatticeofzeros.

Since quasiperiodic meromorphic functions on the torus are determined,
up to a multiplicative constant, by the boundary conditions and the
position and degeneracy of their zeros, one should be able to
write {\it all} sections
\appendixproducts\ in the form
\appendixholoparts\ by mapping the nodal configuration into the set of
complex coefficients $c_\ell.$

Assume then, for a given configuration $\{a_1,\dots,a_k\}$ of
nondegenerate zeros (the degenerate case will be discussed below) that
\eqn\appendixequalholoprod{
\Theta(z) = \sum_{\ell=1}^{k} c_\ell\, \chi^w_\ell(z). }
Then, the $k$ conditions $\Theta(a_i)=0$ imply the following
homogeneous set of equations for the coefficients $c_\ell,$
\eqn\appendixhomoglinear{
\sum_\ell B_{i\ell}\, c_\ell=0,\mtxt{where} B_{i\ell}=\chi_\ell(a_i).}
The equivalence of both parametrisations implies that
the matrix $B=(B_{i\ell})$ has rank $k\!-\!1$
or that its kernel is one-dimensional. Hence the linear
system \appendixhomoglinear\ determines
the coefficients $c_\ell$, up to an overall factor.
The overall constant may be fixed by matching the values of the sections
in both parametrisations at a non-nodal point.
In cases where some zero is degenerate one proceeds
in an anologous way. A node $a$ with
multiplicity $r$ yields $r$ conditions $\Theta(a)=
\Theta^\prime(a)=\dots=\Theta^{(r)}(a)=0$, and the
corresponding rows in $B$ consist of derivatives of
$\chi^w_\ell(z)$ at $a$. \

Hence, any section has the product representation
\eqn\appendixproducts{
\psi(z)=
%= \lambda\; \prod_{i=1}^k \vartheta_1(z-a_i|i). }
\lambda \prod_{i=1}^k \exp{{\pi\over 2}\{(z-a_i)^2 +(1-i)z\}}
\; \theta_1\big( z\!-\!a_i |\,i \big)}
The parameter space for non-trivial holomorphic sections consists
of a nonzero complex constant $\lambda$ and $k$ points on {\af
T}$^2$ subject to the constraint 
\eqn\constraint{ \sum a_i=i\om +\ft{k}{2}(1+i)\;\;\hbox{ mod }\Lambda.} 
As an example consider
$k=2$ and take $w=0$. Then the constraint on the two nodes reads
\eqn\appendixdefinitesum{
a_1 + a_2 = 1+i. }
An unambiguous parametrization of the positions of the zeros
fulfilling this constraint is achieved by picking the node $a_1$
in the region $0 < {\rm Re\ } a_1 < \half$, together with the
segments ${\rm Re\ }a_1 = 0,\ 0\leq{\rm Im\ }a_1 \leq\half$ and
${\rm Re\ }a_1 = \half,\ 0\leq{\rm Im\ }a_1 \leq\half$. The
following figure shows the remaining identifications one needs to
make:
\vfill\eject
 \vskip 13cm {\hskip 2.8cm
\epsfxsize=6cm\epsfbox{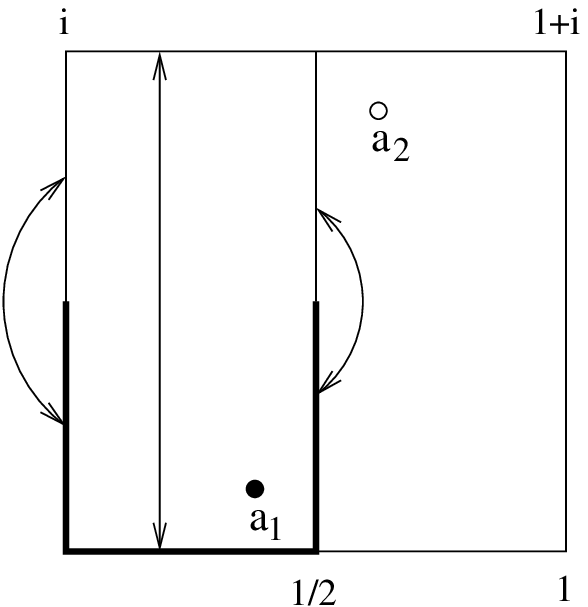}}

\hskip 0.7cm{\vbox{{\bf Figure 4.} The domain ${\cal R}$ for $a_1$
with the necessary identifications.}} \medskip

\noindent This region with identifications is homeomorphic to a
two-sphere $S^2$. The parameter $\lambda$ in \appendixproducts\
contributes a positive constant $|\lambda|$ and a phase arg
$\lambda$. This phase takes values in the fiber of a principal
$U(1)$ bundle over $S^2,$ and to identify it, we need to compute
the first Chern class
\eqn\appendixcherndef{ c_1 (P) = {1\over{4\pi i}}\,
\oint_{\partial{\cal R}} \left( {\rm d}\,\ln\psi - {\rm
d}\,\ln\psi^\ast \right) = {1\over{2\pi i}}\, \oint_{\partial{\cal
R}} {\rm d}\,\ln\psi , }
where ${\cal R}$ is the region defined above.
Since
\eqn\appendixlogpsi{ \ln \psi = \ln \lambda + \ln
\vartheta_1(z-a_1|i) + \ln
\vartheta_1(z-a_2|i)+\hbox{Polynom}\,(z),}
only the theta function associated with the unique zero $a_1\in{\cal R}$
is relevant for the contour integral in \appendixcherndef. Moreover,
from the infinite product expansion for thetas, only a sine factor
contributes:
\eqn\appendixchargesine{ c_1 (P) = {1\over{2\pi i}}\,
\oint_{\partial{\cal R}} {\rm d}\,\ln\vartheta_1(z-a|i) =
{1\over{2\pi i}}\, \oint_{\partial{\cal R}} {\rm
d}\,\ln\sin[\pi(z-a)]=1, }
yielding unit Chern class by the residue theorem. Then the $U(1)$
fibration on $S^2$ is the Hopf bundle $S^3.$

The topologically nontrivial space \R$^4 \setminus \{0\}$ is the
union of all $3$-spheres with radii $|\lambda|\in $\R$_+$. The null
section $\psi(z)=0$ belonging to a singular instanton completes it
to the contractible space \R$^4 \approx $ \C $^2.$

This construction generalizes to arbitrary $k$, since the space of $k$
points on the torus with fixed sum is topologically equivalent to \C$
P^{k-1}.$ The phase arg $\lambda$ defines the sphere $S^{2k-1}$ as a
principal $U(1)$ bundle over this projective space. The space of
sections \R$^{2k} \approx$ \C$^k$ is constructed as before with the
null section and all $S^{2k-1}$ with positive radii. Charge 2 is a
particular case since \C$ P^1 \approx S^2.$

\end
\bye